\documentclass[letterpaper, 10 pt, conference]{ieeeconf}  

\usepackage{graphics} 
\usepackage{epsfig} 
\usepackage{amsmath} 
\usepackage{amssymb}  
\usepackage{hyperref}

\title{\LARGE \bf Probabilistic Planning for Maritime Search and Rescue
}

\author{
Luís Marques$^1$, Jose Javier Escribano Macias$^1$, Panagiotis Angeloudis$^{1*}$\\
$^1$ Imperial College London \quad $^*${\tt\small p.angeloudis@imperial.ac.uk}
}%

\begin{document}

\maketitle
\thispagestyle{empty}
\pagestyle{empty}

\begin{abstract}

Maritime accidents cause thousands of disappearances every year, with migrant crossings being particularly dangerous and under-reported. Current coastal and NGO search and rescue services are unable to provide a timely response, so new technologies such as autonomous UAVs are needed. 
We present a thorough formalization of the maritime search and rescue problem considering its time-critical and probabilistic nature.
Further, we introduce a method for determining the optimal search altitude for any aerial thermal-based detection system, so as to maximize overall mission success.

\end{abstract}

\section{INTRODUCTION}
Drowning is the 3rd leading cause of death worldwide according to a 2021 WHO report \cite{who_death}. Migrant crossings of the Mediterranean Sea lead to massive loss of life with over 26,832 disappearances registered since 2014 \cite{missing_migrants_iom}. The real number of casualties is however much higher, as unnoticed accidents cannot be reported. 

The current search and rescue (SAR) resources of southern European countries are understandably strained and unable to cover such an extensive area in a timely manner. 
Coastguard boats are slower than desired, and helicopters require expensive maintenance and expert operation, while only being capable of rescuing a limited number of people. Thus
unmanned aerial vehicles (UAVs) are being increasingly considered to provide timely first-aid to people at sea \cite{feraruAutonomousUAVbasedSystem2020}. Besides being a cheaper solution than conventional search and rescue units (SRUs), autonomous UAVs are not subject to fatigue and prevent placing SAR operators in danger. UAVs can achieve a lower time-to-target (TTT) than other SRUs due to their shorter setup time and higher cruise speeds. This is crucial for maritime SAR (MSAR) where drowning can be fatal within minutes, and hypothermia can lead to loss of life in as fast as 30 min \cite{survival_hypo}. 

While autonomous UAVs can provide first aid directly, our ongoing work focuses on the application of autonomous UAVs to augment our coverage of Mediterranean Sea crossings (with conventional SRUs or other specialized UAVs providing aid). Improving the coverage along common migrant flow paths would provide us with more accurate and up-to-date accident data, allowing for a better allocation of rescue resources. Most importantly, it would decrease the response time of SRUs and allow us to locate accidents at deep sea that would previously be invisible.

The contributions of this work are as follows: (a) formalization of the MSAR task as a constrained optimization problem, considering its time-critical and probabilistic nature; (b) development of a method for determining the optimal search altitude for any aerial thermal-based detection system.

\section{METHODOLOGY}

\subsection{Problem Definition}
In this Sub-Section, we formalize the overall MSAR task as an optimization problem, allowing us to compare and design search paths for the SRUs. Take $N$ to denote the total number of people to be rescued in a given MSAR scenario. 
If an accident is observed, an estimate of $N$ along with an approximate accident location is available. Ongoing work leverages the definition below, and methods from Sub-Section B, to improve the coverage of migrant paths using multiple autonomous UAVs.

A random variable $N_{saved}(t) \leq N$ is defined to represent the number of people saved up to a given instant $t$. Due to the probabilistic nature of MSAR operations, $N_{saved}(t)$ depends on multiple factors such as the: 1) initial distribution of people along the search area; 2) weather conditions; 3) physical condition of the migrants; 
4) quality of information provided to the SRUs; 
5) initial distance between SRUs and the accident (determines cruise duration);
6) probability of detecting a target in water. The search has duration $\Delta t_{search} = t_f - t_0$, where $t_f$ is the earliest of either complete fuel depletion of the allocated SRUs or safe recovery of all the targets. Let us consider the single-agent case where only one SRU is deployed. The vehicle workspace $\mathcal{W}=\mathbb{R}^3$ and the obstacle region $\mathcal{O}(t) \subset \mathcal{W}$ follow their conventional definitions. Note that $\mathcal{O}$ is a function of time as obstacles need not be stationary. A candidate search trajectory is denoted by $\sigma \in \Sigma$, where $\Sigma$ corresponds to the set of all possible trajectories. 

Thus, we can encode the MSAR task in the following constrained optimization problem with objective function $J$: 
\begin{align}
\min_{\sigma \in \Sigma} \quad & J(\sigma)= \frac{1}{N} \int_{t_0}^{t_f} \left(N-\mathbb{E}[N_{saved}(t)]\right) \ dt
\label{objective-function}
\\
\textrm{s.t.} \quad & \sigma(t) \in \mathcal{W}_{free} =\mathcal{W} \setminus \mathcal{O} , \quad \forall t \in [t_0, t_f]
\label{free-space-constraint}  \\
& F(q,\dot{q},t) = 0 , \quad \forall t \in [t_0, t_f]
\label{dynamic-constraint} \\
& \int_{t_0}^{t_f} P(t) \ dt \leq E_{total}
\label{battery-constraint}
\end{align}

Constraint (\ref{free-space-constraint}) indicates that trajectories should only traverse the free space. Constraint (\ref{dynamic-constraint}) enforces the non-holonomic constraints of boat, helicopter or UAV motion, with $q$ representing the vehicle's configuration. For example, when approximating the dynamics of a fixed-wing UAV via the unicycle model, $q=[x,y,\psi]$ and (\ref{dynamic-constraint}) becomes $\dot{q} - [u\cos \psi, u\sin \psi, \omega]^T = 0$, where $u$ is the forward velocity and $\omega$ the yaw rate. 
Constraint (\ref{battery-constraint}), where $P$ represents power consumption, limits the energy usage of the SRU.
While energy usage is often included in the objective $J$ to maximize path efficiency, it is inconsequential here, as time is of the essence and human lives far outweigh any financial costs.

Before fully defining the objective function (\ref{objective-function}), let us motivate it by considering other intuitive but naive implementations. Great care should always be taken when setting up an optimization problem, especially in life-critical situations. 
One could consider maximizing the coverage of ground area per unit time. This would lead to high search altitudes (for aerial vehicles) and speeds, but ignores how those affect the probability of detecting targets and thus the overall mission success. Alternatively, one could opt to simply maximize $N_{saved}(t_f)$, the total number of people saved at the end of the search. Apart from providing sparse reward signals for learned approaches, this definition would not take into account the time-critical nature of SAR missions.\break
Our proposed $J$ tackles both of these shortfalls. The random variable $N_{saved}(t)$ encapsulates the aleatoric characteristics of the problem, such as how the probability of detecting targets varies with altitude and speed. Its expectation can be empirically determined through simulations (e.g. via Monte Carlo). By integrating the number of people yet to be saved along the mission duration, $J$ incentivizes providing aid to all the targets and doing so as expeditiously as possible. Each person yet to be saved effectively increases the penalty added at each timestep.

\subsection{Optimal Search Altitude}

In this Sub-Section, we develop a method for determining the optimal search altitude for any aerial thermal-based detection system, so as to maximize the probability of saving people at sea.

Compared with other sensing solutions, a long-wave infrared thermal (TIR) camera is uniquely suited for reliably detecting humans at sea. Conventional cameras have limited utility in low/no-light scenarios and are more significantly affected by fog and mist \cite{burkeOptimizingObservingStrategies2019}. Further, most CV research on conventional cameras focuses on images taken close to the target from a ground-based viewpoint, whereas in this application the humans would appear as small blobs in the frame (due to the target's distance, oblique viewpoint and partial submergence). Foreground segmentation would be harder for conventional cameras which are more significantly affected by wave crests and sun shades. Despite having lower spatial and temporal resolution, TIR cameras make foreground segmentation trivial and accessible to onboard computing resources. Maritime backgrounds have far less clutter in the infrared spectrum, since humans appear as bright blobs (as would other homeothermic endotherms on the surface) and the ocean water has uniform temperature.

To help introduce our method, we will perform example calculations using the specifications of the long-wave infrared thermal camera GC3-T \cite{tir_camera}. The GC3-T is a drone TIR camera with IP67 waterproof rating, 50 Hz framerate and 640x512 px resolution. Notwithstanding, the proposed framework is valid for any thermal camera. 
For the purposes of providing aid, we only require target detection (i.e. not recognition or identification),
making relative temperature measurements sufficient. 
TIR sensors must be configured with an appropriate emissivity value to have reasonable readings ($\varepsilon=0.98$ for detecting humans). Since emissivity varies non-linearly as the viewing angle increases above $45^{\circ}$ \cite{nunak_thermal_2015}, we assume the camera is pointing vertically downwards. Further work could consider non-zero tilt angles through active perception, however as the viewing angle increases so does the effect of atmospheric infrared absorption and the spatial ground resolution decreases. Additionally, pointing downwards facilitates calculations, as the ground resolution is constant for all the pixels, and simplifies segmentation, as the horizon and sky stay outside the frame.

\begin{figure}[htbp]
  \centering
  \includegraphics[width=.8\linewidth]{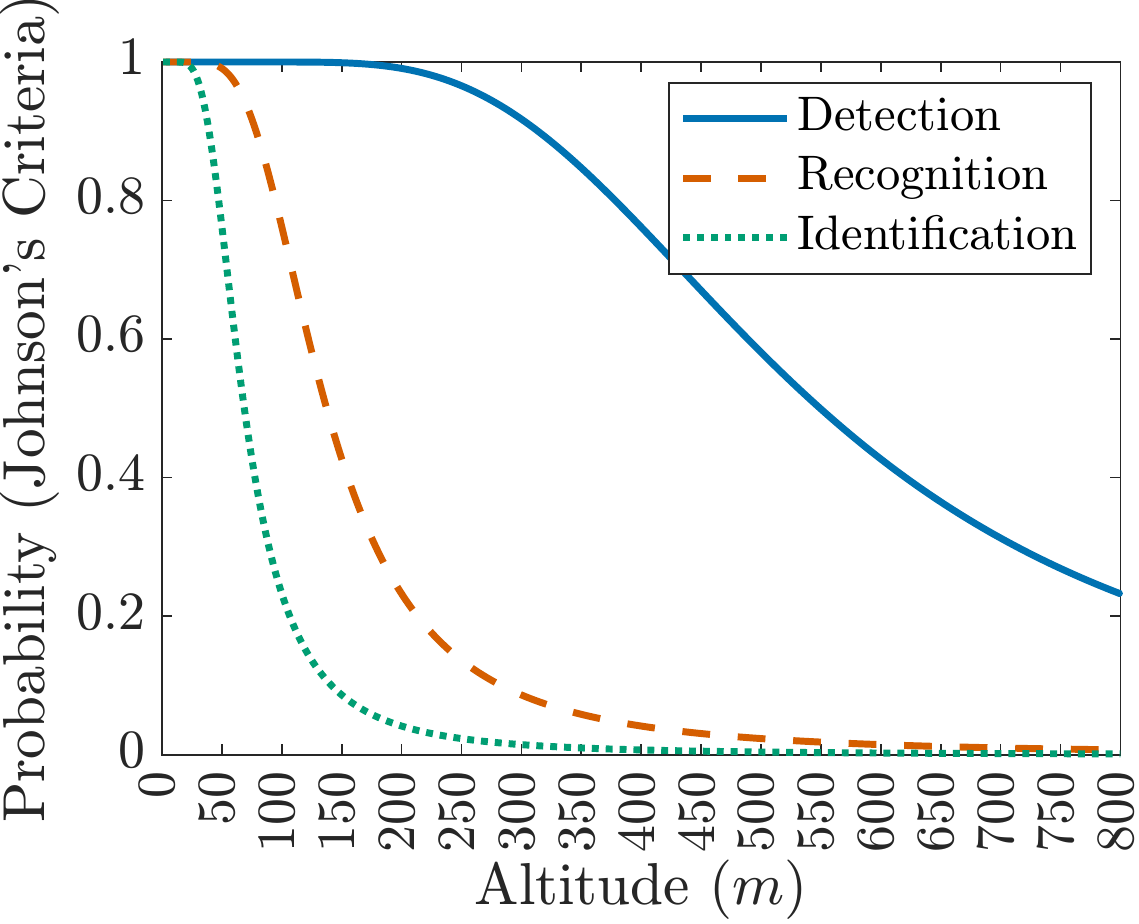}
  \caption{Probability of Detection, Recognition and Identification with altitude according to Johnson's Criteria. The $n_{50}$ values are 0.75, 3.0 and 6.0 respectively. These probabilities were calculated using the GC3-T camera specifications. However, TIR camera resolutions are incredibly standard and cameras with uncooled detectors such as GC3-T all support wider lenses. Telephoto lenses require active cooling, which is significantly more expensive, and cover a significantly smaller ground area, which is a drawback for our application.}
  \label{fig:johnson}
\end{figure}

According to TIR theory, a target should occupy at least 3x3 pixels on the image plane for an accurate thermal reading. However, the MSAR task requires enough detail to correctly classify the object. 
From literature, 5x5 pixels are sufficient for detecting humans using thermal cameras if these are the only expected targets and if false positives are not a major concern \cite{burkeOptimizingObservingStrategies2019,rodinDetectabilityObjectsSea2018}. 
Johnson's Criteria \cite{leachtenauerResolutionRequirementsJohnson2003} provides a theoretical framework for determining the probability of detection in thermal infrared (TIR) systems based on the provided spatial resolution. 
Johnson introduces a factor $n_{50}$ describing the number of cycles required for a 50\% level of performance ($n_{50} = 0.75$ for detection).
The number of cycles $n$ across a target is given by $n = d_c / (2 \cdot \textnormal{GSD})$ where GSD is the ground sample distance and $d_c$ is the target's characteristic dimension.
Succinctly, GSD indicates the ground area seen by a single pixel. 
Thus, a low GSD corresponds to high spatial resolution and a high number of cycles over the target.
We can estimate the probability of detection $P(n)$ for a certain number of cycles $n$ using Johnson's Criteria: 
\begin{equation}
    P(n) = \frac{\left(n/n_{50}\right)^{2.7+0.7\cdot(n/n_{50})}}{1+(n/n_{50})^{0.27+0.7\cdot (n/n_{50})}}
\end{equation}
Considering the average human shoulder width, the likely vertical orientation of a person-in-water and the oblique viewpoint of an aerial thermal image, a target size of 0.5x0.5 m is selected \cite{burkeOptimizingObservingStrategies2019,rodinDetectabilityObjectsSea2018}. Thus $d_c = \sqrt{0.5^2 + 0.5^2} = 0.5$ m.
From camera physics, we can relate GSD to the search altitude by recalling that $\textnormal{GSD}_H = 2R \tan \left(\frac{\textnormal{FOV}_H}{2\cdot \textnormal{PX}_{H}}\right)$ and $
\textnormal{GSD}_V =  \frac{2R}{\cos\phi} \tan \left(\frac{\textnormal{FOV}_V}{2\cdot \textnormal{PX}_{V}}\right) $, where $\textnormal{FOV}$ is the lens' field of view, $\textnormal{PX}$ the sensor size in pixels, $R$ the distance to the target, $\phi$ the viewing angle and $(\cdot)_H$, $(\cdot)_V$ refer to horizontal and vertical dimensions respectively.  
Using these relationships, we can now determine how the probability of detecting a human varies with search altitude. This is shown in Figure \ref{fig:johnson}.

Multiple metrics have been developed in the search theory literature to evaluate search paths \cite{IAMSAR}. We consider the probability of detection (POD), probability of containment (POC) and probability of success (POS) due to their overwhelming popularity, and will use them in our method for determining the optimal flight altitude.

POD indicates the likelihood of detecting a target, given it is within the sensors' range. We have just shown how to determine the POD for any given search height through Johnson's Criteria.

\begin{figure}[htbp]
  \centering
  \includegraphics[width=.9\linewidth]{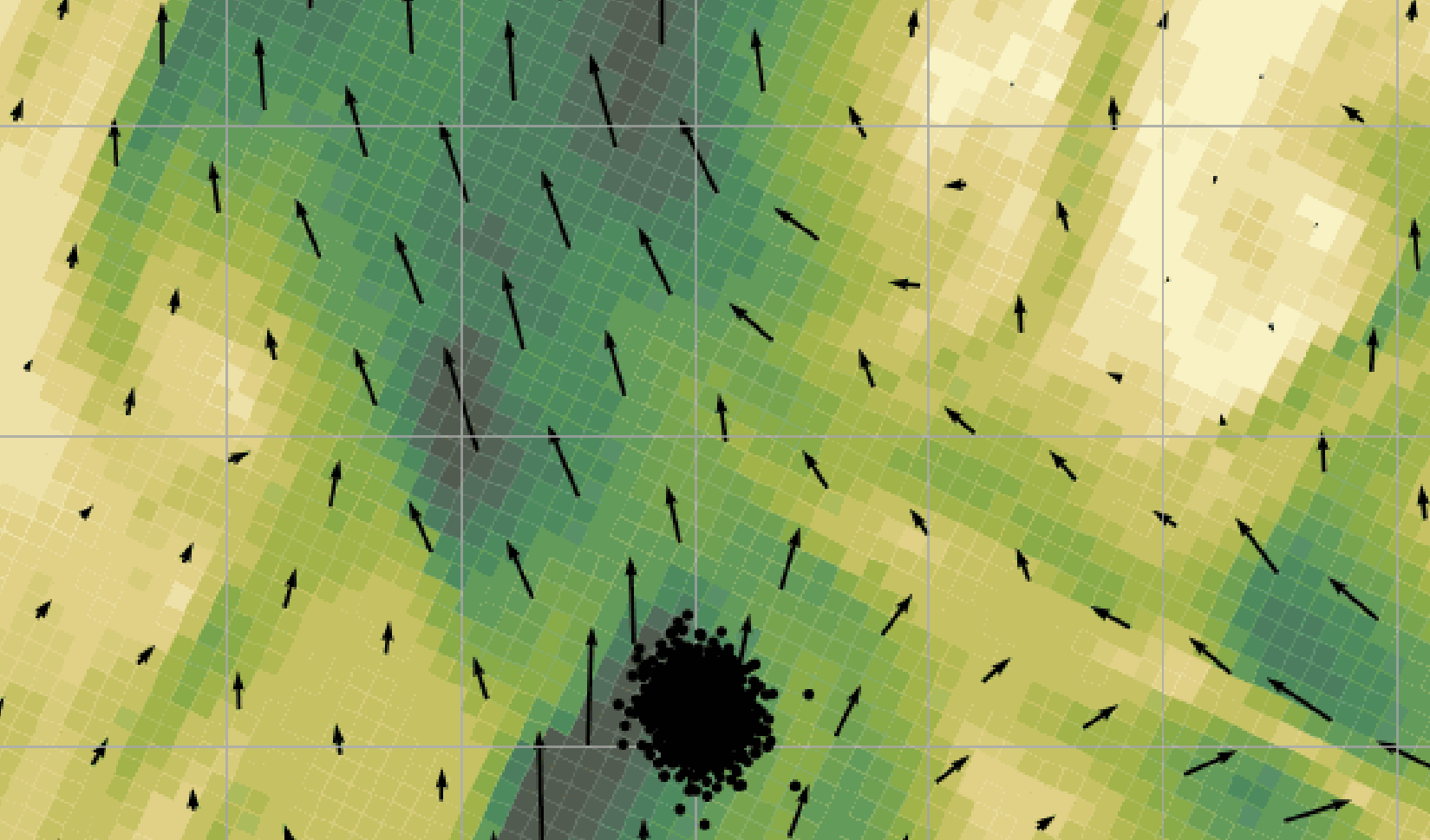}
  \caption{Particle-based drift modelling in OpenDrift. The object type was selected to be a conscious person-in-water in a vertical position. A large number of particles provides  conservativeness at the cost of a more demanding simulation.}
  \label{fig:leeway}
\end{figure}

POC indicates the likelihood of the target being within a given area. Following common practice, we take all targets to be initially located inside the search area. 
This assumption is further supported by our conservative search area estimate detailed below. 
As distance measurements taken from binoculars at sea have a 9\% error at 9 km \cite{kinzeyDistanceMeasurementsUsing2003}, we can conservatively consider the initial accident location provided by the observers to have an 800 m error. 
Taking this initial search area, we ran a 20 min particle-based drift simulation in OpenDrift with 10'000 particles. Using the highest average wind and current speeds of the Mediterranean, the search area increased to 1200x1200 m. A snapshot of the simulation is shown in Figure \ref{fig:leeway}

\break
For a given search altitude, the corresponding field size is given by $\textnormal{FS}_{dir} = \textnormal{GSD}_{dir} \cdot \textnormal{PX}_{dir}$. Assuming a uniform target distribution along the search area, the POC at any given instant is obtained by dividing the ground area observed ($\textnormal{FS}_{H}\cdot \textnormal{FS}_{V}$) by the total search area. A uniform distribution is best suited for when location data is uncertain \cite{IAMSAR}, being more conservative than a Gaussian.

\begin{figure}[htbp]
  \centering
  \includegraphics[width=.8\linewidth]{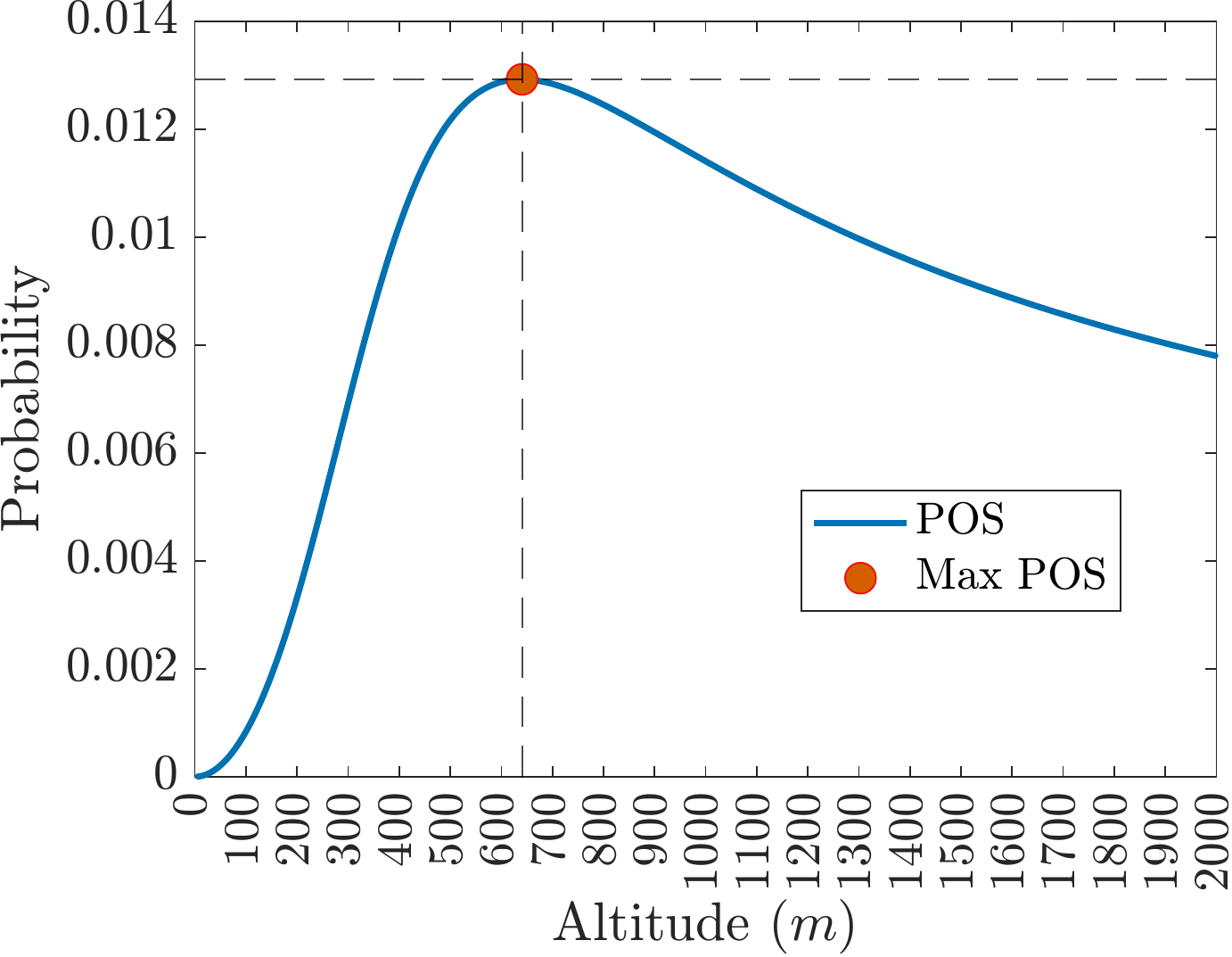}
  \caption{$\textnormal{POS}$ variation with search altitude. The probability values shown are for a given instant and not the whole mission (as the POC is relative to the field size). As height increases so does POC, but POD decreases according to Johnson's Criteria. There is a single height value that maximizes POS.}
  \label{fig:max_pos}
\end{figure}

Finally, $\textnormal{POS} = \textnormal{POC} \cdot \textnormal{POD}$ defines the probability of successfully finding the search object. It is the ultimate measure of search effectiveness. As shown in Figure \ref{fig:max_pos}, we can leverage Johnson's Criteria to determine the unique search altitude that maximizes $\textnormal{POS}$ and thus mission success. 

 The camera framerate places a theoretical upper bound on the search velocity, for a given altitude, to achieve full ground coverage. Practically, search speed is more tightly constrained by aerodynamic and structural considerations.

\section{CONCLUSIONS}
In this work, we formalized the MSAR problem fully, considering its inherently random and time-critical nature. Ongoing work leverages this formulation to compare multiple heuristic search paths across various parameterizations. 

We showed how search altitude affects overall system performance and devised a method for determining the optimal search height for any aerial vehicle using TIR cameras for human detection. The coupling between sensing and planning must be considered when devising an optimal search path that maximizes the probability of survival.
 
We are currently researching how autonomous UAVs can be used to improve our coverage of migrant paths. We are developing methods that build on this work to determine the optimal UAV configuration (payload characteristics, propellant type, etc.), fleet size and deployment type for this application, considering the current NGO and governmental resources. Results will be validated via Monte Carlo simulations on the problem formulation presented here.

\addtolength{\textheight}{-12cm}   

\bibliographystyle{IEEEtran}
\newpage
\bibliography{new_refs}

\end{document}